\documentclass[sigconf]{acmart}
\DeclareMathOperator*{\argmax}{arg\,max}
\usepackage[ruled,linesnumbered]{algorithm2e}
\usepackage{setspace}

\usepackage{graphicx}
\usepackage{xcolor}
\usepackage{hyperref}
\usepackage{booktabs} 
\usepackage{subfigure}
\usepackage{color}
\usepackage{enumitem}
\usepackage{multirow}
\usepackage{multicol}
\usepackage{ulem}
\usepackage[nomargin,inline,marginclue,draft]{fixme}
\usepackage{balance}
\usepackage{verbatim}
\usepackage{diagbox}
\usepackage{changepage}
\usepackage{xcolor}

\newlist{todolist}{itemize}{2}
\setlist[todolist]{label=$\square$}
\usepackage{pifont}

\newcommand{\rev}[1]{#1}

\newtheorem{mdefinition}{Definition}

\newcommand{\para}[1]{{\vspace{4pt} \bf \noindent #1 \hspace{0pt}}}

\newlength\savedwidth

\begin{document}

\title{Efficient Data-specific Model Search \\ for Collaborative Filtering}

\author{Chen Gao$^{1}$, Quanming Yao$^{1,2}$, Depeng Jin$^{1}$, Yong Li$^{1}$}
\affiliation{\institution{$^{1}$Beijing National Research Center for Information Science and Technology (BNRist),\\
		Department of Electronic Engineering, Tsinghua University}
}
\affiliation{\institution{$^{2}$4Paradigm Inc.}}

\begin{abstract}
Collaborative filtering (CF), 
as a fundamental approach for recommender systems, 
is usually built on the latent factor model with learnable parameters to predict users' preferences towards items. 
However, 
designing a proper CF model for a given data is not easy, 
since the properties of datasets are highly diverse.
In this paper, 
motivated by the recent advances in automated machine learning (AutoML), 
we propose to design a data-specific CF model by AutoML techniques.
The key here is a new framework that unifies state-of-the-art (SOTA) CF methods and splits them into disjoint stages of input encoding, 
embedding function, interaction function, and prediction function.
We further develop an easy-to-use, robust, and efficient search strategy, 
which utilizes random search and a performance predictor for efficient searching within the above framework.
In this way, we can combinatorially generalize data-specific CF models, which have not been visited in the literature, from SOTA ones.
Extensive experiments on five real-world datasets demonstrate that our method can consistently outperform SOTA ones for various CF tasks. 
Further experiments verify the rationality of the proposed framework and the efficiency of the search strategy. 
The searched CF models can also provide insights for exploring more effective methods in the future.
\end{abstract}

\maketitle

\footnotetext[1]{The first two authors contributed equally to this research.}

\section{Introduction}

Collaborative filtering (CF) is the most widely used approach for recommender systems~\cite{koren2008factorization,sarwar2001item,he2017neural}. 
Its idea is based on the generally-accepted assumption that users who have similar historical preferences tend to have the same preferences in the future. 
Early memory-based approaches~\cite{sarwar2001item,linden2003amazon} that directly calculate history's similarity are seldom used now due to their inferior recommendation performance.
Recently,
model-based approaches~\cite{koren2008factorization,he2017neural,kabbur2013fism},
which builds latent factor models to predict users' preferences with learnable parameters, have become the mainstream solution.

However,
the datasets used in different CF tasks have various properties, 
such as form, scale, distribution, etc.,
since they arise from a wide range of applications.
For example, there are two representative forms of CF datasets,
implicit and explicit,
and datasets may be diverse in the scale (large or small) and distribution (dense or sparse). 
On the one hand, matrix factorization methods, such as naive 
MF~\cite{koren2008factorization} or FISM~\cite{kabbur2013fism}, 
are easy-to-train but cannot capture complicated user-item interactions due to their limited capacity~\cite{yao2019searching}.
On the other hand, neural models may achieve better performance when data is sufficient but 
may not perform well for relatively small datasets. 
For example, NCF~\cite{he2017neural}, the widely-acknowledged state-of-the-art neural CF model, can outperform MF methods steadily with a relatively large dataset. 
A recent study~\cite{dacrema2019we} has demonstrated that these neural models do not necessarily achieve better performance on some evaluation datasets.
Therefore, designing CF model should be a \textit{data-specific} problem.

\begin{figure}[t]
	\includegraphics[width=1.0\linewidth]{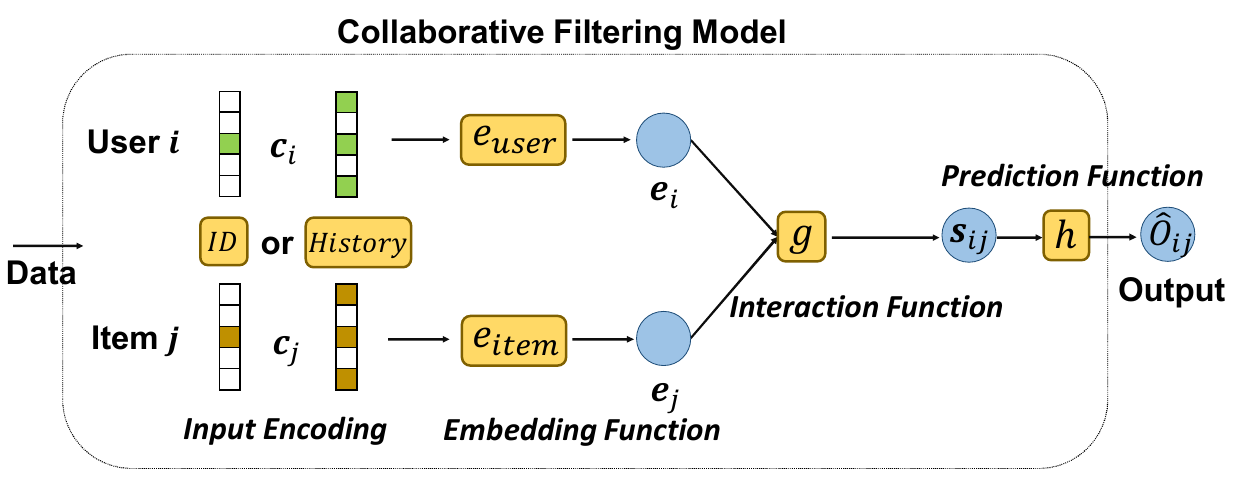}
	
	\caption{A unified framework of CF models, which contain four stages: input encoding, embedding function, interaction function and prediction function.}
	\label{fig:framework_method}
\end{figure}

Recently, Automated Machine Learning (AutoML)~\cite{hutter2019automatedck,quanming2018auto} achieved great success in various machine learning tasks,
e.g.,
neural architecture search~\cite{zoph2017neural,pham2018efficient,liu2018darts,yao2019differentiable,li2019random} and
hyper-parameter optimization~\cite{bergstra2012random,li2017hyperband,feurer2019auto,liu2021learnable},
we propose to search data-specific CF models by AutoML techniques.
However, there are two-fold challenges for AutoML-based CF problem:
\begin{itemize}[leftmargin=*]
	\item 
	First, it is challenging to design a proper search space that can not only cover existing human-crafted methods but also explore more models that have superior performance for a given dataset.
	
	\item Second, since the 
	datasets' properties are quite diverse,  
	the search strategy is required to be efficient, robust and ease-to-use.
\end{itemize}

To solve the first challenge, we propose a general framework that unifies state-of-the-art CF models.
In general, most machine learning models consist of three key stages, data encoding, representation learning, and prediction function.
CF models also follow this paradigm but with one significant extra stage, 
user-item matching, 
which is required after the representation learning. 
To be more specific, 
CF's key is to match users with proper items, and this additional stage matches user representations with item representations.
Thus, 
the general framework for CF models consists of four stages 
(as in Figure~\ref{fig:framework_method}).
First, the model encodes the input of user and item. 
Second, the model uses dense vectors to represent them. 
Third, the model matches user representations and item representations. 
Finally, a prediction function obtains the score.

To solve the second challenge, we develop a robust, 
efficient,
and easy-to-use search strategy,
which is based on
the random search and a performance predictor.
To be specific, with random search from the above-designed search space and a performance predictor to fast evaluate the searched models, our approach can efficiently find powerful models in the space for different datasets.

Our contributions can be summarized as follows.
\begin{itemize}[leftmargin=*]
	\item We approach the problem of CF task with the new perspective of automated machine learning. Since it is data-specific, we design AutoML based method to search proper CF model for a given dataset.
	
	\item We propose a general search space based on a standard four-stage paradigm of human-designed models for CF, including input coding, embedding function, interaction function, and prediction function. The proposed search space can not only cover most of the human-crafted models but also find more models not still explored by human experts.
	We develop an effective search strategy, which is easy-to-use and robust, to search data-specific CF models.
	
	\item Extensive experiments on five real-world datasets demonstrate that our proposed approach can efficiently search powerful models that achieve the best recommendation performance, with improvements of 1.65\% - 22.66\%.
	Analysis of the searched models provides insightful guidance to help human expert design models for CF.
	Further empirical results verify the effectiveness in search space and strategy design.
\end{itemize}

\begin{table*}[ht]
	\small
	\centering
	\caption{Representative Human-designed CF Models. (Single models refer to those with one group of user and item embeddings; for fused models, there are multiple embedding matrices for user and item, bringing more learnable parameters. Here we have $u_i = \texttt{MAT}(\mathrm{ID}(i))$ and $v_j = \texttt{MAT}(\mathrm{ID}(j))$, and we use $u_i$ and $v_j$ for better presentation.)}\label{tab:human_designed_models}
	\setlength\tabcolsep{5pt}
	\begin{tabular}{c|c|c|c|c|c|c|c|c}
		\toprule
		       \bf Category         &                      \bf  Name                       &                                             \bf Model Formulation                                              & \bf  User Enc. $c_i$ & \bf  Emb. $e_{\text{user}}$ & \bf  Item Enc. $c_j$ & \bf  Emb. $e_{\text{item}}$ & \bf  Interacton $g$ & \bf  Prediction $h$ \\ \midrule
		\multirow{8}{*}{\bf Single} &                          MF                          &                                       $\mathbf{u}_i \cdot \mathbf{v}_j$.                                       &     \texttt{ID}      &        \texttt{MAT}         &     \texttt{ID}      &        \texttt{MAT}         &  \texttt{multiply}  &    \texttt{SUM}     \\ \cmidrule{2-9}
		                            &              FISM~\cite{kabbur2013fism}              &                              $ \texttt{MAT}({\mathbf{r}_i}) \cdot \mathbf{v}_j $                               &   \texttt{History}   &         \texttt{ID}         &     \texttt{MAT}     &        \texttt{MAT}         &  \texttt{multiply}  &    \texttt{SUM}     \\ \cmidrule{2-9}
		                            &               GMF~\cite{he2017neural}                &                                $\texttt{VEC}(\mathbf{u}_i \odot \mathbf{v}_j)$                                 &     \texttt{ID}      &        \texttt{MAT}         &     \texttt{ID}      &        \texttt{MAT}         &  \texttt{multiply}  &    \texttt{VEC}     \\ \cmidrule{2-9}
		                            &               MLP~\cite{he2017neural}                &                                  $\texttt{MLP} (\mathbf{u}_i ; \mathbf{v}_j)$                                  &     \texttt{ID}      &        \texttt{MAT}         &     \texttt{ID}      &        \texttt{MAT}         &   \texttt{concat}   &    \texttt{MLP}     \\ \cmidrule{2-9}
		                            &          CMF~\cite{hsieh2017collaborative}           &                                      $|| \mathbf{u}_i - \mathbf{v}_j ||$                                       &     \texttt{ID}      &        \texttt{MAT}         &     \texttt{ID}      &        \texttt{MAT}         &   \texttt{minus}    &    \texttt{Norm}    \\ \cmidrule{2-9}
		                            &                DMF~\cite{xue2017deep}                &                         $\texttt{MLP}(\mathbf{r}_i) \cdot  \texttt{MLP}(\mathbf{r}_j)$                         &   \texttt{History}   &        \texttt{MLP}         &   \texttt{History}   &        \texttt{MLP}         &  \texttt{multiply}  &    \texttt{SUM}     \\ \cmidrule{2-9}
		                            &            JNCF-Cat~\cite{chen2019joint}             &                    $\texttt{MLP}(\texttt{MLP}(\mathbf{r}_i) ; \texttt{MLP}(\mathbf{r}_j))$                     &   \texttt{History}   &        \texttt{MLP}         &   \texttt{History}   &        \texttt{MLP}         &   \texttt{concat}   &    \texttt{MLP}     \\ \cmidrule{2-9}
		                            &            JNCF-Dot~\cite{chen2019joint}             &                  $\texttt{MLP}(\texttt{MLP}(\mathbf{r}_i) \odot \texttt{MLP}(\mathbf{r}_j))$                   &   \texttt{History}   &        \texttt{MLP}         &   \texttt{History}   &        \texttt{MLP}         &  \texttt{multiply}  &    \texttt{MLP}     \\ \cmidrule{1-9}
		\multirow{7}{*}{\bf Fused}  &      \multirow{2}{*}{NeuMF~\cite{he2017neural}}      & \multirow{2}{*}{$ \texttt{VEC}(\mathbf{u}_i \odot \mathbf{v}_j) +  \texttt{MLP}(\mathbf{u}_i ; \mathbf{v}_j)$} &     \texttt{ID}      &        \texttt{MAT}         &     \texttt{ID}      &        \texttt{MAT}         &  \texttt{multiply}  &    \texttt{VEC}     \\ \cmidrule{4-9}
		                            &                                                      &                                                                                                                &     \texttt{ID}      &        \texttt{MAT}         &     \texttt{ID}      &        \texttt{MAT}         &   \texttt{concat}   &    \texttt{MLP}     \\ \cmidrule{2-9}
		                            & \multirow{2}{*}{SVD++~\cite{koren2008factorization}} &     \multirow{2}{*}{ $\mathbf{u}_i \cdot \mathbf{v}_j + \texttt{MAT}({\mathbf{r}_i}) \cdot \mathbf{v}_j $}     &     \texttt{ID}      &        \texttt{MAT}         &     \texttt{ID}      &        \texttt{MAT}         &  \texttt{multiply}  &    \texttt{SUM}     \\ \cmidrule{4-9}
		                            &                                                      &                                                                                                                &   \texttt{History}   &        \texttt{MAT}         &     \texttt{ID}      &        \texttt{MAT}         &  \texttt{multiply}  &    \texttt{SUM}     \\ \cmidrule{2-9}
		                            &      \multirow{4}{*}{DELF~\cite{cheng2018delf}}      &                                {$\texttt{MLP}(\mathbf{u}_i  ; \mathbf{v}_j) $}                                 &     \texttt{ID}      &        \texttt{MAT}         &     \texttt{ID}      &        \texttt{MAT}         &   \texttt{concat}   &    \texttt{MLP}     \\ \cmidrule{4-9}
		                            &                                                      &                         $ + \texttt{MLP}(\mathbf{u}_i ; \texttt{MLP}(\mathbf{r}_j)) $                          &     \texttt{ID}      &        \texttt{MAT}         &   \texttt{History}   &        \texttt{MAT}         &   \texttt{concat}   &    \texttt{MLP}     \\ \cmidrule{4-9}
		                            &                                                      &                          $+ \texttt{MLP}(\texttt{MLP}(\mathbf{r}_i) ; \mathbf{v}_j )$                          &   \texttt{History}   &        \texttt{MAT}         &     \texttt{ID}      &        \texttt{MAT}         &   \texttt{concat}   &    \texttt{MLP}     \\ \cmidrule{4-9}
		                            &                                                      &                   $+ \texttt{MLP}(\texttt{MLP}(\mathbf{r}_i) ; \texttt{MLP}(\mathbf{r}_j))$                    &   \texttt{History}   &        \texttt{MAT}         &   \texttt{History}   &        \texttt{MAT}         &   \texttt{concat}   &    \texttt{MLP}     \\ \bottomrule
	\end{tabular}
\end{table*}

\subsubsection*{Notation}
The dataset utilized in the CF task is a user-item interaction matrix $\mathbf{Y}^{M\times N}$ with $M$ users and $N$ items. 
Each entry of this matrix can be binary or ratings (from 1 to 5, for example). 
The objective of CF is to estimate the missing values of the matrix. In other words, 
the output of a CF model is $\hat{\mathbf{O}}_{ij}$ for user $u$ and item $j$.
\texttt{MAT}$(\cdot)$ represents to multiply the multi-hot representation with a matrix and then apply to mean pooling; since the embedding lookup for ID and embedding lookup for history with mean pooling are quite similar, both of them are denoted with \texttt{MAT}. 
For the final prediction, \texttt{SUM} denotes multiply with an all-one vector (\textit{i.e.}, sum), \texttt{VEC} denotes multiply with a learnable vector, and \texttt{MLP} stills denotes feeding a multi-layer perception. We use $(\cdot;\cdot)$ to denote a concatenation operation and $ \odot$ to denote the element-wise product.

\section{Related Work}\label{sec::relatedwork}

\subsection{Collaborative Filtering (CF)}\label{sec::relatedwork1}
Collaborative filtering (CF)~\cite{sarwar2001item,he2017neural}, learning user preferences based on user-item interaction, is the most fundamental solution for recommender systems. Memory-based CF~\cite{sarwar2001item} is the earliest CF approach that calculates the similarity of the whole interaction history. 
Recently, model-based CF
which builds latent factor models with learnable parameters and predicts users' preference score towards items directly, becomes more popular.
We summarize the representative CF models in Table~\ref{tab:human_designed_models}.
MF~\textit{et al.}~\cite{koren2008factorization} was proposed to project user ID and item ID with real-value vectors, representing user interests and item features, respectively. Then MF uses the inner product to predict the preference score with user and item embeddings. CMF~\textit{et al.}~\cite{hsieh2017collaborative} was proposed that uses minus operation to replace MF's inner product.

Beside IDs, user history and item history\footnote{History means the interaction history.} can also be used to represent the encoding, which has been demonstrated effective in SVD++~\cite{koren2008factorization}, FISM~\cite{kabbur2013fism} and SLIM~\cite{ning2011slim}.
These early progresses only use explicit rating data, such as user-movie rating scores. However, in modern recommender systems, there is plenty of implicit feedback data, such as the purchase or click records. Therefore, the item-ranking task, which takes the implicit data as input and predicts a ranking list, is more important.
BPR~\cite{rendle2009bpr} proposed a pairwise learning manner to optimize the MF model on implicit data.  
Inspired by the big success of the deep neural networks in computer vision and natural language processing~\cite{lecun2015deep}, NCF~\cite{he2017neural} applied neural networks to building CF models. NCF used a multi-layer perceptron (MLP) as the interaction function, taking user and item embedding vectors as input and predict the preferences score. Similarly, DeepMF~\textit{et al.}~\cite{xue2017deep} proposed a deep matrix factorization model that replaced user/item ID with user/item history and reserved the inner product operation in MF.
Furthermore, JNCF~\textit{et al.}~\cite{chen2019joint} extended NCF by using user/item history to replace user/item ID as the input encoding. \rev{The authors proposed two variants, JNCF-Cat and JNCF-Dot, with different functions to match user embeddings and item embeddings.}
There is a recent work, SIF~\cite{yao2019searching}, that adopts the one-shot architecture search for adaptive interaction function in the CF model.

Recently, researchers found that the neural model does not necessarily perform better than shallow models~\cite{dacrema2019we}.
In other words, whether a model is suitable or not largely depends on the dataset.
Since CF tasks are various, on totally different datasets with diverse properties, \textit{the best model} could be different. In other words,  designing models for CF is a \textit{data-specific} problem. 


\subsection{Automated Machine Learning (AutoML)}\label{sec::relatedwork2}

Automated machine learning  (AutoML)~\cite{hutter2019automatedck,quanming2018auto} refers to a type of method that can
self-adapt machine learning models to various tasks. 
Recently, 
AutoML has achieved a great success in designing the state-of-the-art model in various learning
applications such as image classification/segmentation~\cite{zoph2017neural,tan2019efficientnet,liu2019auto}, 
natural language modeling~\cite{so2019evolved}, 
and knowledge graph embedding~\cite{zhang2019autokge}.
A general framework of AutoML consists of three parts as follows:
\begin{itemize}[leftmargin=*]
	\item \textit{Search Space}: it defines the set of all possible models, which be finite or infinite. 
	Search space is designed with human expertise on specific applications. The search space should be carefully chosen since an overlarge space brings challenges to model search will a too-small space cannot cover powerful models. 
	
	\item \textit{Search Algorithm}: 
	it refers to the algorithm 
	that guides the searching process in the space.
	Popular choices are random search~\cite{bergstra2012random}, 
	evolutionary algorithms~\cite{xie2017genetic}, Bayesian optimization~\cite{bergstra2011algorithms}
	and reinforcement learning~\cite{zoph2017neural}.
	Among them,
	while being simple,
	random search is a strong baseline in many applications~\cite{bergstra2011algorithms,li2019random}.
	
	\item \textit{Model Evaluator}: it measures the performance of a candidate, i.e., a model, 
	in the space,
	which is also usually the most expensive part in AutoML.
	The most basic evaluator is direct model training.
	Early-stop~\cite{maron1994hoeffding,hutter2011sequential} is a popular
	trick to cut-down training cost of bad candidates. Performance predictor~\cite{eggensperger2015efficient,liu2018progressive,zhang2019autokge}
	extended it by avoid training of unpromising candidate via an additional model.
\end{itemize}

Since search space requires a lot of domain-specific knowledge,
it is very diverse for current AutoML applications, 
e.g., 
from shallow models~\cite{thornton2013auto,feurer2019auto,zhang2019autokge} to deep neural architectures~\cite{pham2018efficient,zoph2017neural}. 
Search algorithm and 
model evaluator highly couple with each other.
Specifically,
the search algorithm focuses on generating the next  
candidates in the search space,
while the evaluator offers feedbacks to the algorithm.
Thus,
the computational cost of
AutoML depends on both the number of candidates generated
(by the algorithm) and the evaluating time of candidates (by the evaluator).

\section{Problem Definition}\label{sec::problemdef}

As mentioned in the introduction, designing CF models is data-specific. 
Based on the proposed framework,
we propose to use AutoML to efficiently search CF model for the given dataset.

\subsection{A Unified Framework of CF Model}\label{sec::cf_model}

As mentioned in the introduction,
CF models also follows the paradigm of general machine learning, 
i.e., data encoding, representation learning, and prediction function,
while one significant additional stage, user-item matching, is required after the representation learning. 
To be more specific, the key to CF is to match users with proper items, 
and this additional stage matches user representations with item representations.
In general, the four stages derived from existing CF models are in Figure~\ref{fig:framework_method} and details are as follows:

\begin{enumerate}[leftmargin=*]
	\item \textbf{Input encoding} represents the input of user and item. The two main choices are ID-based one-hot encoding or history-based multi-hot encoding, which represents a user with her historically interacted items (\textit{i.e.} one row in user-item interaction matrix), as follows,
	\begin{equation}
	\begin{aligned}
	&\mathbf{c}_i = \mathrm{ID}(i) \,\, \mathrm{or} \,\,  \mathbf{r}_i, \\
	&\mathbf{c}_j = \mathrm{ID}(j) \,\, \mathrm{or} \,\, \mathbf{r}_j, \\
	\end{aligned}
	\end{equation}
	
	where $\mathbf{r}_i$ ($\mathbf{r}_j$) is $i$-th row ($j$-th column) of the user-item matrix.
	
	\item \textbf{Embedding function} projects the input encoding to dense real-value embeddings, which represent the user interests or item features, denoted as $\mathbf{e}_i $ and $\mathbf{e}_j$ respectively, as follows,
	\begin{equation}
	\mathbf{e}_i = e_{\text{user}}(\mathbf{c}_i), 
	\quad
	\mathbf{e}_j = e_{\text{item}}(\mathbf{c}_j), 
	\end{equation}
	
	\item \textbf{Interaction function} matches the user and item embeddings in the latent space for further prediction, as follows,
	\begin{equation}
	\mathbf{s}_{ij} = g(\mathbf{e}_i, \mathbf{e}_j).
	\end{equation}
	
	\item \textbf{Prediction function} converts the output of interaction function into predicted value of user-item preferences, as follows,
	\begin{equation}
	\hat{\mathbf{O}}_{ij} = h(\mathbf{s}_{ij}).
	\end{equation}
\end{enumerate}

Here we present some empirical results to reveal each stage's significant impact on recommendation performance 
(see full results in Table~\ref{tab::rating_perform} and Table~\ref{tab::ranking_perform}). 
FISM~\cite{kabbur2013fism} that utilizes user history as input encoding performs better than MF in some tasks (item ranking), while fails in others. 
The similar finding can also be observed in the interaction function by comparing CMF~\cite{hsieh2017collaborative} and MF and in the prediction function by comparing GMF~\cite{he2017neural} and MF.
The impact of four stages' operation choices is in two folds. First, a bad choice of any stage may make the whole recommendation model not work at all.
Second, whether the choice of each stage is good or not is closely related to the specific datasets, as datasets' different properties require different operations.
As a result, the design of CF model is a \textit{data-specific} problem involving the operation selection of four stages.

\subsection{Formulating as an AutoML Problem}

Recently, researchers found that neural models do not necessarily perform better than shallow models~\cite{dacrema2019we}, 
which makes us rethink the CF task.
Since CF tasks are various, with totally different datasets and performance measurement, 
\textit{the best model} should be different. 
There is no one-for-all model that can always achieve good performance on all CF tasks.
In other words, model design for CF is a \textit{data-specific} problem. 
As discussed above, since different datasets' properties have a significant impact on the performance of CF models, 
we are motivated to form the problem of designing new and better CF models as a search problem, as follows:

\begin{mdefinition}[Automated Model Search for CF (AutoCF)] Let $f^{*}$ denote the desired CF model. Then the CF search problem can be formulated as:
	\begin{align}
		f^* 
		& = \max\nolimits_{f \in \mathcal{F}}\mathcal{M}(f(\mathbf{P}^{*}), \mathcal{S}_{\text{val}}),
		\label{eq:autocf1}
		\\
		\text{\;s.t.\;} 
		&
		\mathbf{P}^* 
		= \argmax\nolimits_\mathbf{P} \mathcal{M}(f(\mathbf{P}), \mathcal{S}_{\text{tra}}),
	\end{align}
	where $\mathcal{F}$ contains all possible choices of $f$, $\mathcal{S}_{\text{val}}$ and $\mathcal{S}_{\text{tra}}$ denote the training and validation datasets, $\mathbf{P}$ denotes the learnable parameters of the CF model $f$, and $\mathcal{M}$ denotes the performance measurement.
\end{mdefinition}

To solve the AutoCF problem well, there are two important aspects.
First, the search space should cover various operations, which make sure for capturing various characteristics of different datasets.
Second, the search strategy should be both robust and efficient that can meet the requirements of both accuracy and efficiency in real-world applications.

Compared with previous AutoML solutions~\cite{tan2019efficientnet,liu2018darts,liu2019auto}, 
the AutoCF problem is quite different since the search space and strategy are closely related to the domain of personalized recommender systems.
First, both shallow and deep models are demonstrated to be effective by existing works~\cite{he2017neural, dacrema2019we}. 
Second, the task's settings in real-world applications are diverse, which requires an efficient search strategy to search powerful models for a given dataset.

\section{Our Approach}\label{sec::method}

Following a commonly-recognized paradigm of AutoML, solving the AutoCF problem should consider three parts, search space, search algorithm, and model evaluator, as mentioned in Section~\ref{sec::relatedwork2}. Our AutoCF approach is featured as follows:
\begin{itemize}[leftmargin=*]
	\item A general \textbf{search space}, i.e., $\mathcal{F}$, 
	that not only covers effective state-of-the-art solutions, but also explores new models.
	\item An easy-to-use and robust \textbf{search strategy} including 1) a random search algorithm that is simple yet effective and 2) a performance predictor based model evaluator to help fast evaluate the searched models.
\end{itemize}

Recently, SIF~\cite{yao2019searching} propose a one-shot architecture search algorithm for finding good interaction functions in rating prediction tasks.
In this work, we consider the problem from a more general perspective, besides the interaction function, with a full model search including input encoding, embedding function, interaction function, and prediction function, on both implicit and explicit feedback data. In addition, the search algorithm in~\cite{yao2019searching} cannot be applied to searching full CF models for our problem since it can only search operations with the same input/output.


\subsection{Search Space Design}\label{sec::search_space}
To solve the challenge of search space design, we unify the state-of-the-art CF models in a general framework, which follows a four-stage paradigm. 

\subsubsection{Operation Selection}
The used operations for the search space are illustrated in Table~\ref{tab::operations}.

\rev{
	\para{Input Encoding} 
	There are no extra features in CF tasks, such as user profiles or item attributes; thus, for representing user and items, an intuitive manner is \texttt{ID}, using the one-hot ID and maintaining two embedding matrices. Another manner is \texttt{History}, to represent a user with his/her multi-hot history (\textit{i.e.} interacted items), and items can be represented similarly.
	That is, encoding refers to how to represent the user and item in the CF task.
}

\para{Embedding Function}
Embedding function that projects the high-dimensional input encoding to low-dimension embeddings is closely related to input encoding operation. With \texttt{ID} input encoding, the only compatible embedding function for user/item is \textit{user/item embedding matrix ID-look-up}, denoted as \texttt{MAT}.
With \texttt{history} input encoding, there are two kinds of compatible embedding functions.
First, 
the user/item embedding function of \textit{item/user embedding matrix ID-lookup and mean pooling} is used in some representative models such as FISM~\cite{kabbur2013fism} and SVD++~\cite{koren2008factorization}, denoted as \texttt{MAT}. The item embedding function can be built similarly.
Second, a multi-layer perceptron can also be used to convert multi-hot history into dense vectors directly, denoted as \texttt{MLP}. Here for the architecture of MLP, we follow the setting in the state-of-the-art neural model, NeuMF~\cite{he2017neural}.

\para{Interaction Function}
Interaction function connect two sides, \textit{i.e., user and item}, for subsequent prediction. A frequently-used interaction function is to use element-wise product, \texttt{multiply}, to generate a combined vector. 
Similar operation consists of \texttt{minus}, \texttt{max}, \texttt{min} and \texttt{concat}.

\para{Prediction Function}
The prediction function converts the output of the interaction function to a predicted score.
A simple yet effective operation is \textit{summation}, denoted as \texttt{SUM}. \textit{Inner product with a weight vector} can be used to assign weights to a different dimension, denoted as \texttt{VEC}.
The \textit{multi-layer perceptron} can also be used for more complicated prediction, denoted as \texttt{MLP}.

\subsubsection{Full Model Encoding}
For a whole model, we need to define an encoding to represent the above choices of each stage.
Since one-hot representation has been demonstrated effective by previous works~\cite{zhang2019autokge,liu2018progressive}, 
we adopt it to represent each stage's operation. To make the space more general, we also consider embedding dimension choice from a set $\mathcal{S}_{\text{dim}}$ and optimizer' learning rate from a set $\mathcal{S}_{\text{lr}}$ as part of the encoding. Then size of the whole space is $135*|\mathcal{S}_{\text{dim}}| * |\mathcal{S}_{\text{lr}}|$.
For example, 
$(\{0,1\}$, 
$\{1,0\}$, 
$\{1,0\}$, 
$\{0,1\}$, 
$\{0,0,1,0,0\}$, 
$\{1,0,0\}$, 
$\{0,0,1,0\}$, 
$\{0,1,0,0\})$ represents the model with ({\texttt{History}, \texttt{ID}, \texttt{MAT}, \texttt{MLP}, \texttt{min}, \texttt{SUM}}) learned with $3$-th embedding dimension size and $2$-th learning rate choice, if the size of $\mathcal{S}_{\text{dim}}$  and $\mathcal{S}_{\text{lr}}$  are both set to $4$.  This corresponds the whole paradigm in Figure~\ref{fig:framework_method} and operation choice in Table~\ref{tab::operations}.

\begin{table}[ht]
	\small
	\centering
	\caption{The operations we use for the search space.}\label{tab::operations}
	\begin{tabular}{c|c|c}
		\hline
		\bf Stage    & \multicolumn{2}{c}{\bf Operations} \\ \hline
		\multirow{2}{*}{\bf Input Encoding }  &  User &\texttt{ID}, \texttt{History}  \\ \cline{2-3}
		& Item &\texttt{ID},\, \texttt{History}  \\ \hline
		\multirow{2}{*}{\bf Embedding Function}   &  User &\texttt{Mat},\, \texttt{MLP}   \\ \cline{2-3}
		& Item &\texttt{Mat},\, \texttt{MLP}   \\ \hline
		\bf  Interaction Function   & \multicolumn{2}{c}{ \texttt{MUL},\, \texttt{MINUS},\, \texttt{MIN},\, \texttt{MAX},\, \texttt{CONCAT}}   \\ \hline
		\bf Prediction Function &\multicolumn{2}{c}{ \texttt{SUM},\, \texttt{VEC}, \, \texttt{MLP} }\\ \hline
	\end{tabular}
\end{table}

\subsection{Search Strategy}

To solve the challenge of search efficiency, we propose an easy-to-use and robust search strategy that combines random search with performance predictor.
The whole search strategy is present in Algorithm \ref{algo::algo1}.
Given the defined search space, 
our problem turns to design efficient and proper search strategy to find models with excellent performance. 
Note that our problem is also suffering the problem of discrete search space, 
and thus random search,
which is a strong baseline in many AutoML applications~\cite{li2019random,zoph2017neural,sciuto2020evaluating}, 
can be a simple and effective solution.

However,
the evaluation in random search is expensive.
To make the search process efficient, 
we propose to combine random search with a performance predictor that can quickly evaluate the performance of searched models.
\subsection{Experimental Settings}
\begin{table*}[t]
	\small
	\caption{Comparison on Rating-Prediction Task.}\label{tab::rating_perform}
	\setlength\tabcolsep{4pt}
	\begin{tabular}{cccccccccc}
		\toprule
		\multicolumn{2}{c}{\bf Dataset}      & \multicolumn{2}{c}{\bf MovieLens-100K} & \multicolumn{2}{c}{\bf MovieLens-1M} & \multicolumn{2}{c}{\bf Yelp} & \multicolumn{2}{c}{\bf Amazon-Book} \\ \midrule
		\multicolumn{2}{c}{\bf Metric}       &   \bf RMSE    &     \bf        MAE     & \bf    RMSE &    \bf         MAE     & \bf   RMSE  & \bf       MAE  &  \bf  RMSE  &    \bf         MAE    \\ \midrule
		\multirow{8}{*}{\bf Single} &     MF      &    0.8964     &         0.4906         &   0.8152    &         0.7061         &   1.1031    &     0.8204     &   1.0661    &        0.9910         \\
		\multirow{8}{*}{\bf Model}  &    FISM     &    0.9212     &         0.4645         &   1.0379    &         0.9516         &   1.0746    &     1.2510     &   1.0265    &        1.1273         \\
		&     GMF     &    0.8872     &         0.4993         &   1.2125    &         1.0259         &   1.0413    &     0.7473     &   1.0249    &        0.9148         \\
		&     MLP     &    0.8632     &         0.4665         &   0.8135    &         0.6588         &   0.9218    &     0.6598     &   0.8803    &        0.7565         \\ 
		&     DMF     &    0.5118     &         0.4682         &   0.8043    &         0.6577         &   0.9051    &     0.6587     &   0.8458    &        0.7563         \\
		&  JNCF-Dot   &   {0.4703}    &        {0.4146}        &   0.7946    &        {0.6101}        &   0.9084    &     0.6553     &  {0.8469}   &       {0.7620}        \\
		&  JNCF-Cat   &    0.4587     &         0.4124         &  {0.7990}   &         0.6041         &   0.9029    &    {0.6559}    &   0.8454    &        0.7534         \\
		&     CMF     &    0.7151     &         0.4232         &   1.2688    &         0.8722         &   1.4401    &     1.0088     &   1.3474    &        1.2708         \\ 
		&   AutoCF    &    0.4479     &         0.3876         &  {0.7933}   &         0.5911         &  {0.9012}   &     0.6410     &  {0.8431}   &        0.7351         \\ \midrule
		& Improvement & \bf    2.24\% &       \bf 6.01\%       & \bf 1.65\%  &       \bf 2.15\%       & \bf 2.18\%  &   \bf 4.56\%   & \bf 3.65\%  &      \bf 3.35\%       \\ \midrule
		\multirow{5}{*}{\bf Fused}  &    SVD++    &    0.8809     &         0.4244         &   0.8295    &         0.6708         &   1.0710    &     0.7449     &   0.9945    &        0.8334         \\
		\multirow{5}{*}{\bf Model}  &    NeuMF    &    0.7923     &         0.4590         &   0.7909    &         0.7366         &   0.9910    &     0.7673     &   0.9259    &        0.8729         \\
		&    DELF     &    0.5357     &         0.3912         &   0.7904    &         0.8036         &   0.9807    &     0.8553     &   0.9106    &        0.8538         \\
		& SinBestFuse &    0.4215     &         0.3942         &   0.7827    &         0.6038         &   0.8994    &     0.6429     &   0.8368    &        0.7356         \\
		&   AutoCF    &    0.4075     &         0.3516         &   0.7744    &         0.5861         &   0.8805    &     0.6233     &   0.8235    &        0.7155         \\ \midrule
		& Improvement & \bf  10.96\%  &      \bf 11.08\%       & \bf 10.99\% &      \bf 11.21\%       & \bf 10.98\% &  \bf 11.30\%   & \bf 11.01\% &      \bf 11.18\%      \\ \bottomrule
	\end{tabular}
\end{table*}

\begin{algorithm}[ht]
	\normalem 
	\caption{AutoCF: Automated Model Search for CF}
	\label{algo::algo1}
	
	\SetKwInOut{Input}{input}
	\SetKwInOut{Output}{output}
	\SetKwRepeat{Do}{do}{while}

	\Input{Search space $\mathcal{F}$, a learnable predictor  $\mathcal{P}$, performance measurement $\mathcal{M}$, a empty set $\mathcal{H}$, size of training batch for predictor $K_1$ and $K_2$, training data $\mathcal{S}$.}
	
	Initialize the predictor $\mathcal{P}$ with random parameters\;

	\Do{not meet stop criteria}{
		Randomly select a $(K_1+K_2)$-size model set $\mathcal{F}^{b}$ from $\mathcal{F}$\;
		Generate one-hot encodings $\mathbf{x}_o$ to represent models in $\mathcal{F}^{b}$\;
		Estimate the performance of models in $\mathcal{F}^{b}$ with $\mathcal{P}$ \;
		Choose top-$K_1$ model sets $\mathcal{F}^{t}$ to train with $\mathbf{\mathcal{S}}$\;
		Evaluate the trained models in $\mathcal{F}^{t}$ with $\mathbf{\mathcal{M}}$  \;
		Update the set of evaluated-model $\mathcal{H} \leftarrow \mathcal{H} \cup 
		\{ (f, \mathcal{M}(f)) |f \in \mathcal{F}^{t} \} $ \;
		Update $\mathcal{P}$ with records in $\mathcal{H}$ via loss function in (\ref{eqn::mlp_predictor_loss}).
	}
	\Return desired CF models in $\mathcal{H}$.
\end{algorithm}

\subsubsection{Predictor Design}

However, the bottleneck in our problem is the evaluation of models.
Inspired by previous AutoML works~\cite{eggensperger2015efficient,liu2018progressive,zhang2019autokge}, 
we propose a predictor to distinguish good and bad models. 
Predictor, as mentioned in Section~\ref{sec::relatedwork2}, 
is a model that can estimate the performance given each model's features without training.

It is worth mentioning that many previous AutoML tasks utilize 
weight-sharing techniques~\cite{liu2018darts,pham2018efficient} to accelerate model evaluator. 
For example, the searched different CNN models can share parameters in the a cell without re-training in~\cite{pham2018efficient}. 
However, in our AutoCF problem, the search space cannot be constructed to a \textit{super net} (a directed acyclic graph (DAG) of which a path denotes a single model)~\cite{pham2018efficient} due to the disjoint stages of CF models, which makes the weight-sharing technique cannot be applied.

Predictor aims to predict performance given a model, and thus we need to represent each one in the space with unique encoding. 
As introduce in Section~\ref{sec::search_space},
the encoding is denoted as $\{\mathbf{x}_o\}$ with $o$ from $1$ to $4$.
As discussed in Section~\ref{sec::relatedwork2}, multi-layer perceptron and tree-based models such as random forest are the two mainstream models for predictor.
In our problem, the multi-layer perceptron supports parameter updates with gradient descent, which is more compatible with random search and has a stronger ability to learn from complex data. 
Thus, we use a multi-layer perceptron to predict the model's performance based on the encodings, as follows,
\begin{equation}\label{eqn::mlp_predictor}
\mathcal{P}(\mathbf{x}_o) = \text{\tt MLP}( \text{\tt Concat}(\mathbf{x}_o)),
\end{equation}
where 
the predicted recommendation performance can be in various forms, such as rating prediction accuracy or ranking recall.

\subsubsection{Predictor Training}
For finding good models efficiently, it is not necessary to predict the absolute value of a specific metric since it is more meaningful to distinguish the better model with several candidates.
Therefore, we train the predictor with pairwise loss~\cite{rendle2009bpr}. To be specific, the predictor's objective is to rank pairs with the right order based on evaluated models.
The loss function is formulated as follows,
\begin{equation}\label{eqn::mlp_predictor_loss}
L_{\mathcal{P}} = \sum\nolimits_{(\mathbf{x}_+, \mathbf{x}_-)\in O} - 
\log\big( \sigma(\mathcal{P}(\mathbf{x}_+)-\mathcal{P}(\mathbf{x}_-)) \big),
\end{equation}
where $O$ are built pairs that consists of two searched architectures, and $\sigma$ denotes the \textit{sigmoid} function. $\mathbf{x}_+$ and $\mathbf{x}_-$ denotes the model encoding of the better architecture and the worse one, respectively.


In short, a standard job procedure is first sampling models from the space and then utilizing the predictor to choose top models for real evaluation, and lastly, using evaluation results to update the predictor. This job procedure is repeated under random search until finding powerful models for a given task.

When the pipeline stops (stop criteria can be a hyper-parameter and chosen for different purposes), we obtain a valid predictor that can predict all models' performance with high accuracy and find effective models from the search space.

\section{Experiments and Evaluations}\label{sec::exp}

The mainstream recommendation tasks can be divided into two categories, rating prediction, and item ranking. Based on these two categories, we conduct experiments to answer the following research questions.
\begin{itemize}[leftmargin=*]
	\item Can our method find data-specific CF models that outperform human-design ones on benchmark datasets?
	
	\item What insights can we see from searched models? 
	
	\item How about our designed search space compared with other possible search space? How can the designed predictor speed up search algorithms? 
\end{itemize}

\subsubsection{Datasets}

In the experiments, we use MovieLens-100K, MovieLens-1M, Yelp, and Amazon-Book for the rating-prediction task, and use MovieLens-100K, MovieLens-1M, Yelp, and Pinterest for the item-ranking task. The details, including statistics, of datasets are shown in Appendix~\ref{sec::appen::dataset}.

\subsubsection{Metrics and Loss Function}
As the objective of two tasks is different, 
metrics for evaluating recommendation models are also different. For the task of rating prediction, we use the following two widely used standard metrics of RMSE and MAE~\cite{chai2014root}.

While for the task of item ranking, the mainstream evaluation metrics can be divided into two kinds, recall-ratio based metrics such as Recall and Hit Ratio and rank-position based metrics such as NDCG and MRR~\cite{manning2008introduction}. Here we choose two most widely used metrics, Recall and NDCG.

Similarly, the choice of the loss function is also different for two kinds of tasks. For rating prediction CF tasks, we adopt the widely-used utilized loss function of RMSE loss~\cite{koren2008factorization}; while for item ranking CF tasks, we adopt the BPR loss~\cite{rendle2009bpr}, the state-of-the-art loss function for optimizing recommendation models.

\subsubsection{Hyper-parameters}

Besides the above-mentioned search space, there are also two hyper-parameters for each model: optimizer and negative sampler.
For both two tasks, we choose Adam~\cite{kingma2014adam} optimizer. For item ranking tasks, we need to sample some unobserved items as negative feedback, of which the choice of the negative sampler is another hyper-parameter. For our experiments, we use the most acceptable random negative sampler~\cite{he2017neural, kabbur2013fism, rendle2009bpr} that randomly chooses one unobserved item as a negative sample for each observed user-item feedback.

%


\begin{table*}[t]
	\small
	\caption{Comparison on Item-Ranking Task.}\label{tab::ranking_perform}
	\setlength\tabcolsep{2pt}
	\begin{tabular}{cccccccccc}
		\toprule
		\multicolumn{2}{c}{\bf Dataset}   & \multicolumn{2}{c}{\bf MovieLens-100K} & \multicolumn{2}{c}{\bf MovieLens-1M} & \multicolumn{2}{c}{\bf Yelp} & \multicolumn{2}{c}{\bf Pinterest} \\ \midrule
		\multicolumn{2}{c}{\bf Metric}       &\bf  Recall@20 &        \bf NDCG@20         &\bf  Recall@20 &   \bf     NDCG@20        &\bf  Recall@20 &  \bf    NDCG@20      &\bf  Recall@20 &   \bf    NDCG@20      \\ \midrule
		\multirow{8}{*}{\bf Single}   &       MF       &  0.1073   &         0.1345         & 0.0990   &        0.0589        &  0.0870   &      0.0486      &  0.0965   &      0.0645       \\
		\multirow{8}{*}{\bf Model} &      FISM      &  0.1572   &         0.1422         &  0.1097   &        0.0627        &  0.0928   &      0.0516      &  0.1022   &      0.0703       \\
		&      GMF       & 0.1047   &         0.1412         &  0.1092   &        0.0619        &  0.0939   &      0.0499      &  0.1011   &      0.0691       \\
		&      MLP       &  0.1358  &         0.1221         &  0.0929   &        0.0542        &  0.0807   &      0.0430      &  0.0868   &      0.0602       \\
		&      DMF       &  0.1833   &         0.1213         &  0.0940   &        0.0352        &  0.0800   &      0.0428      &  0.0876   &      0.0602       \\
		&    JNCF-Dot    & 0.2067   &         0.1448         &  0.1255   &        0.0640        &  0.0954   &      0.0503      &  0.1027   &      0.0705       \\
		&    JNCF-Cat    & 0.1875   &         0.1454         &  0.1221  &        0.0642        &  0.0961   &      0.0517      &  0.1045   &      0.0705       \\
		&      CMF       &  0.1450   &         0.1091         &  0.0830   &        0.0489        &  0.0707   &      0.0384      &  0.0778   &      0.0523       \\
		&      AutoCF      &  0.2327   &         0.1755         &  0.1346   &        0.0785        &  0.1155   &      0.0633      &  0.1255   &      0.0862       \\ \midrule
		&  Improvement   &\bf    21.64\%&\bf 21.20\%&\bf 21.26\%&\bf 22.66\%&\bf 21.07\%&\bf 22.44\%&\bf 20.10\%&\bf 22.27\%        \\ \midrule
		\multirow{5}{*}{\bf Fused}  &     SVD++      &  0.1774   &         0.1340         &  0.1028   &        0.0599        &  0.0873   &      0.0470      &  0.0949   &      0.0654       \\
		\multirow{5}{*}{\bf Model}    &     NeuMF      &  0.2114   &         0.1583         &  0.1215   &        0.0695        &  0.1042   &      0.0567      &  0.1124   &      0.0783       \\
		&      DELF      &  0.1931   &         0.1453         &  0.1121   &        0.0642        &  0.0952   &      0.0511      &  0.1038   &      0.0711       \\
		& SinBestFuse &  0.2585   &         0.1934         &  0.1479   &        0.0864        &  0.1264   &      0.0697      &  0.1390   &      0.0940       \\
		&      AutoCF      &  0.2904   &         0.2190         &  0.1677   &        0.0973        &  0.1425   &      0.0777      &  0.1554   &      0.1065       \\ \midrule
		&  Improvement   &  \bf   12.34\%&\bf 13.24\%&\bf 13.39\%&\bf 12.62\%&\bf 12.74\%&\bf 11.48\%&\bf 11.80\%&\bf 13.30\%        \\ \bottomrule
	\end{tabular}
\end{table*}

\begin{table*}[t]
	\small
		\caption{Top 3 searched models of each dataset.
			\texttt{H} represents \texttt{History}.}\label{tab:top}
	\centering
	\setlength\tabcolsep{2pt}
	\begin{tabular}{c|c|cc|cc|cc}
		\hline
		      \bf Task        & \bf  Dataset &                                          \bf  Top1                                           & \bf  Perf. &                                          \bf   Top2                                          & \bf  Perf. &                                          \bf  Top3                                           & \bf  Perf. \\ \hline
		                      &   ML-100K    &   $\big<$\texttt{H},\texttt{H},\texttt{MLP},\texttt{MLP},\texttt{max},\texttt{MLP}$\big>$    &   0.4479   &   $\big<$\texttt{H},\texttt{H},\texttt{MLP},\texttt{MLP},\texttt{max},\texttt{VEC}$\big>$    &   0.4512   &   $\big<$\texttt{H},\texttt{H},\texttt{MLP},\texttt{MLP},\texttt{min},\texttt{MLP}$\big>$    &   0.4520   \\
		\bf Rating Prediction &    ML-1M     & $\big<$\texttt{H},\texttt{H},\texttt{MLP},\texttt{MLP},\texttt{multiply},\texttt{VEC}$\big>$ &   0.7933   &   $\big<$\texttt{H},\texttt{H},\texttt{MAT},\texttt{MLP},\texttt{min},\texttt{MLP}$\big>$    &   0.7939   &   $\big<$\texttt{H},\texttt{H},\texttt{MLP},\texttt{MLP},\texttt{max},\texttt{MLP}$\big>$    &   0.7945   \\
		       (RMSE)         &     Yelp     &   $\big<$\texttt{H},\texttt{H},\texttt{MLP},\texttt{MLP},\texttt{plus},\texttt{SUM}$\big>$   &   0.9012   &  $\big<$\texttt{H},\texttt{H},\texttt{MLP},\texttt{MLP},\texttt{concat},\texttt{SUM}$\big>$  &   0.9015   &   $\big<$\texttt{H},\texttt{H},\texttt{MLP},\texttt{MAT},\texttt{min},\texttt{SUM}$\big>$    &   0.9021   \\
		                      & Amazon-book  &   $\big<$\texttt{ID},\texttt{H},\texttt{MLP},\texttt{MLP},\texttt{max},\texttt{VEC}$\big>$   &   0.8431   &   $\big<$\texttt{ID},\texttt{H},\texttt{MLP},\texttt{MLP},\texttt{min},\texttt{VEC}$\big>$   &   0.8432   & $\big<$\texttt{ID},\texttt{H},\texttt{MAT},\texttt{MLP},\texttt{concat},\texttt{VEC}$\big>$  &   0.8447   \\ \hline
		                      &   ML-100K    & $\big<$\texttt{H},\texttt{H},\texttt{MAT},\texttt{MAT},\texttt{multiply},\texttt{VEC}$\big>$ &   0.2327   & $\big<$\texttt{H},\texttt{H},\texttt{MAT},\texttt{MAT},\texttt{multiply},\texttt{MLP}$\big>$ &   0.2197   &   $\big<$\texttt{H},\texttt{H},\texttt{MLP},\texttt{MLP},\texttt{min},\texttt{VEC}$\big>$    &   0.2149   \\
		  \bf Item Ranking    &    ML-1M     & $\big<$\texttt{H},\texttt{H},\texttt{MAT},\texttt{MAT},\texttt{multiply},\texttt{VEC}$\big>$ &   0.1346   &   $\big<$\texttt{H},\texttt{H},\texttt{MLP},\texttt{MLP},\texttt{min},\texttt{MLP}$\big>$    &   0.1282   & $\big<$\texttt{H},\texttt{H},\texttt{MLP},\texttt{MLP},\texttt{multiply},\texttt{MLP}$\big>$ &   0.1255   \\
		     (Recall@20)      &     Yelp     & $\big<$\texttt{H},\texttt{H},\texttt{MLP},\texttt{MLP},\texttt{multiply},\texttt{VEC}$\big>$ &   0.1155   & $\big<$\texttt{H},\texttt{H},\texttt{MAT},\texttt{MAT},\texttt{multiply},\texttt{VEC}$\big>$ &   0.1091   &   $\big<$\texttt{H},\texttt{H},\texttt{MLP},\texttt{MLP},\texttt{min},\texttt{VEC}$\big>$    &   0.1077   \\
		                      &  Pinterest   & $\big<$\texttt{H},\texttt{H},\texttt{MAT},\texttt{MAT},\texttt{multiply},\texttt{SUM}$\big>$ &   0.1255   & $\big<$\texttt{H},\texttt{H},\texttt{MAT},\texttt{MAT},\texttt{multiply},\texttt{MLP}$\big>$ &   0.1210   &   $\big<$\texttt{H},\texttt{H},\texttt{MLP},\texttt{MLP},\texttt{min},\texttt{VEC}$\big>$    &   0.1169   \\ \hline
	\end{tabular}
\end{table*}

\subsection{Benchmark Performance Comparison}

\subsubsection{Single Model Comparison}
According to the definition in Section 2, \textit{single models} only use one group of embeddings for users and items. Here we choose the nine mainstream single models introduced in Section~\ref{sec::relatedwork1} for comparison. Note that all methods can be adapted to two tasks by transforming loss function, even if they may be designed for one task, \textit{e.g.}, NCF~\cite{he2017neural} is originally proposed for item ranking tasks.\footnote{In fact, a lot of existing CF models~\cite{he2017neural, kabbur2013fism} has claimed in the original paper that they can be adapted to different recommendation tasks by changing loss function.}

We present the obtained RMSE and MAE performance in Table~\ref{tab::rating_perform}. From the results, we can observe that our proposed method searches a model achieving the best performance on all metrics across the four datasets. The improvement is steady for different datasets, and our proposed method can outperform the best baseline by 1.65\% to 6.01\%, which demonstrate the our approach can effectively self-adapt to different tasks to find powerful models.
Another interest observation is that JNCF-Cat, as the best baseline, outperforms all other baselines. This can be explained that use rating-history can introduce more useful signals for embeddings. In other words, the rating-history can be considered as a kind of feature input. For this, we will conduct more analysis in Section 5.3.

For the item ranking tasks, we use the full-ranking protocol~\cite{rendle2019evaluation} to calculate the Recall@K and NDCG@K. Precisely, for each test item, we first calculate the interaction probability of all unobserved items (excepted for items sampled for pair-wise learning); then, we rank the predicted scores for metric calculation. Here we set K to 20, following common settings of many recent works\footnote{According to experimental results in existing works on CF~\cite{he2017neural, chen2019joint}, for one given metric such as Recall or NDCG, the selection of top-K has minor impacts on the partial order of model performance. Our approach can search model given both metric and top-K, and we leave studies on other top-K choices as a future work.}~\cite{wang2019kgat,wang2019neural}.
We present the performance on four utilized datasets in Table~\ref{tab::ranking_perform}. \rev{It is worth mentioning that there are differences between our results and original papers for two reasons. First, the data splitting and pre-processing always have some differences. Second, the ranking metrics are highly relevant to the length of the list and the choices of top-K. In our experiments, we choose the full-ranking, which is a more reasonable choice, which makes our results have large difference compared with those sampling-ranking papers~\cite{he2017neural}.}We can observe that our searched models can achieve the best performance compared with all other baselines. Note that our proposed method can outperform the best baseline by 20.10\% to 22.66\%.


\subsubsection{Fused Model Comparison}

Some researches also find the strong power of 
model fusion in CF. 
In the original paper of NCF, the authors propose a model named NeuMF that fusion a generalized MF and MLP together.
Cheng~\textit{et al.}~\cite{cheng2018delf} proposed a dual model named DELF that fusions four extensions of MLP as a whole model.


Although those fused models can achieve better performance, their larger parameter size and training/inferring cost are always a concern. For better comparison, we also fuse the top-2 best single models, comparing with the recent advances in recommender systems, which also fuses two single models: SVD++ fuses MF and FISM, NeuMF fuses GMF and MLP, and DELF fuses several MLP-based models. We also choose the top-2 baselines in single-model experiments and fuse then as a competitive baseline method, denoted as SinBestFuse.

We present the performance of rating prediction and item ranking in Table 3 and 4. We can easily find that our fused model can achieve the best performance, improving the best baseline by 10.96\%-13.39\%. 
In short, with model fusion that can take advantage of different good models in the space, the recommendation performance can be further improved.

\subsection{Case Study: Data-specific Models}

For the searched models, we present the top ones for each task on all datasets in Table~\ref{tab:top}.
In general, the searched models of different tasks are quite different, which further validates the necessity of considering CF as a data-specific problem.
We can find that for these top models of each task, there are always similar input encoding. 
Another finding is that the impact of the prediction function is the smallest. On the ML-100K datasets, the top-2 models even only differ in prediction function.
In order to present the impact of each stage more clearly, we conduct the analysis of each one as follows,

\begin{itemize}[leftmargin=*]
	
	\item \textbf{Input Encoding.} 
	Most top-models have a common CF prediction signal, $\{\texttt{History}, \texttt{History}\}$. However, for the Amazon-Book dataset, models with $\{\texttt{ID}, \texttt{History}\}$ as the raw input of embedding function can also achieve good performance. This can be explained that user-side records in Amazon-book are more sparse than other datasets.  
	
	\item \textbf{Embedding Function.}
	For different tasks, the choice of embedding function has big differences. We can observe that \texttt{MLP} tend to perform better in the rating prediction task, while \texttt{MAT} can achieve good performance in item ranking tasks. This verifies that our method can distinguish the difference and select suitable operations for different tasks.
	\item \textbf{Interaction Function.}
	For the interaction function, we can first find that our method can choose the proper one for two different tasks on different datasets. Furthermore, for MovieLens-100K where exists a larger popularity bias, interaction functions can learn non-personalized bias such as \texttt{min} and \texttt{max} can achieve good performance.
	\item \textbf{Prediction Function.} For the choice of prediction function, sparse datasets tend to choose simple \texttt{VEC} or \texttt{SUM}, which have few parameters and is easy to learn.
\end{itemize}

\begin{table*}[t!]
		\small
	\centering
	\caption{Comparison of search algorithms with/without predictor.}\label{tab::search_algo}
	\begin{tabular}{c|c|ccccc}
		\hline
		&             & \multicolumn{4}{c}{\bf   Number of Evaluated Models.} \\ \hline
		\bf Task                        &\bf  Dataset     & Rand    & Reinforce    & Rand(P) &Reinforce(P)  & Reduction    \\ \hline
		\multirow{4}{*}{\bf Rating Prediction} & ML-100K     & 1,420                & 1,450        & 150   & 170   & 89.44\%          \\
		& ML-1M       & 1,450                & 1,250        & 370  & 360    & 74.48\%              \\ 
		& Yelp        & 1,310                & 1,180        & 450   & 470  & 65.65\%        \\ 
		& Amazon-book & 1,420                & 1,390        & 520  & 530    & 63.38\%           \\ \hline
		\multirow{4}{*}{\bf Item Ranking}    & ML-100K     & 1,250                & 1,170        & 360   & 400   & 71.20\%                      \\ 
		& ML-1M       & 1,510                & 1,470        & 330   & 350  & 78.15\%              \\ 
		& Yelp        & 1,240                & 1,240        & 510     & 500 & 58.87\%             \\ 
		& Pinterest   & 1,570                & 1,570        & 500  & 530    & 68.15\%         \\ \hline            
	\end{tabular}
\end{table*}

In summary, the best models are not the same for different datasets but sharing some powerful choice of operations, such as input encoding. 
These results, along with analysis, can inspire human experts to design more powerful models.

\subsection{Ablation Study}
\label{sec::search_algo}

\subsubsection{Plausibility of the Search Space}
To further validate the effectiveness of our designed search space, 
we propose to extensively search the architectures for the state-of-the-art model,  
NeuMF~\cite{he2017neural}, with reinforcement learning-based controller~\cite{zoph2017neural}.
As described in~\cite{he2017neural}, the architecture of multi-layer perceptrons can have many implementations. 
With an extensive architecture search, 
we present the performance comparison of the best-searched model in Table~\ref{tab::search_space}. 
From these results, we can find our search space can always obtain better architectures compared with NeuMF's space for different CF tasks. 
Under our designed search space, the best method can improve NeuMF+NASNet by 1.11\%-35.00\% for different tasks. 
As the discussion in Section~\ref{sec::relatedwork1}, it was found~\cite{dacrema2019we} deep models do not necessarily perform better performance compared with shallow models. Our empirical results further validate this observation.
In summary, we conclude that our designed search space is more proper for CF tasks and cover more powerful models, 
based on the observed performance improvement.

\begin{table}[ht]
	\small
	\caption{Performance comparison of different search space.}\label{tab::search_space}
	\begin{tabular}{c|c|ccc}
		\hline
		               &              & \multicolumn{3}{c}{\bf  Best Performance in Search Space} \\ \hline
		   \bf Task    & \bf  Dataset & NeuMF+NASNet & AutoCF &              Improv.              \\ \hline
		  \bf Rating   &   ML-100K    &    0.6754    & 0.4075 &              35.00\%              \\
		\bf Prediction &    ML-1M     &    0.7831    & 0.7744 &              1.11\%               \\
		    (RMSE)     &     Yelp     &    0.9530    & 0.8805 &              7.60\%               \\
		               & Amazon-book  &    0.8640    & 0.8235 &              4.69\%               \\ \hline
		   \bf Item    &   ML-100K    &    0.2327    & 0.2904 &              24.79\%              \\
		 \bf  Ranking  &    ML-1M     &    0.1399    & 0.1677 &              14.36\%              \\
		 (Recall@20)   &     Yelp     &    0.1121    & 0.1425 &              27.18\%              \\
		               &  Pinterest   &    0.1218    & 0.1554 &              27.58\%              \\ \hline
	\end{tabular}
\end{table}

\subsubsection{Acceleration from the Predictor}
In this work, we propose a predictor based search strategy that can find effective models in space.
Here we replace the random search with Reinforce~\cite{zoph2017neural}.
{Reinforce}~\cite{zoph2017neural} is the most widely used search strategy in the neural architecture search. 
To adapt it to our problem, we design a multi-layer perceptron as the controller of which the output is the operation choice. 
By using the recommendation performance of the sampled models as a reward, we adopt the policy gradient (PG) to update the controller. Here we combine it with our proposed performance predictor and named it as Reinforce(P).

\begin{figure}[ht]
	\centering
	\subfigure{\includegraphics[width=0.6\columnwidth]{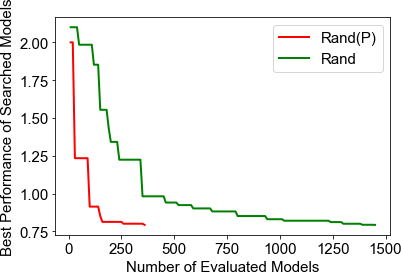}}
\caption{Test RMSE curve with or without predictor.}
	\label{fig::seach_efficiency}
\end{figure}

We present the cost of evaluation of our search strategy in Table~\ref{tab::search_algo}.  
The details of rating prediction task on ML-1M is shown in 
Figure~\ref{fig::seach_efficiency}.
The cost is measured by the number of evaluated models before finding the best model. 
We can find that compared with two baseline methods, our proposed search strategy is quite efficient, improving the best baseline by 58.9\%-89.4\%.
We can observe that the Reinforce search algorithm has quite a similar search efficiency with the random search without predictor. This further validates our discussion in Section~\ref{sec::relatedwork2}: for those tasks of which the search space is discrete, the problem on evaluation (for the lower-level problem) may be more critical than the choice of search algorithm (for the lower-level problem).
In summary, our search algorithm is efficient and has high application value.

\section{Conclusion and Future Work}\label{sec::conclusion}
In this work, we consider the model design in CF as a data-specific problem and approach this problem with automated machine learning. With an elaborately designed search space and efficient search strategy, our AutoCF approach can generate better models than human-designed baselines with various datasets in multiple real-world applications. Our proposed predictor is demonstrated to be more efficient compared with the existing search strategy. Further experiments verify the rationality of our designed search space. The case study on searched powerful models provide insights for developing CF models. First, some operation choices always achieve good performance, such as using history as the input encoding; second, there are some operations mainly depending on the datasets.

For the future work, we plan to transfer the searched top models in our utilized datasets to datasets with an industrial scale.  This will make our approach an effective solution to handle the CF model design in very large datasets.
Another important future work is to expand the definition of CF to cover more possible operations, including embedding propagation operation~\cite{wang2019neural}.
We also plan to consider more extensive recommendation tasks besides CF, such as content-based recommendation~\cite{rendle2010factorization}.
\section{Acknowledgments}\label{sec::acknowledgments}
\sloppy{}
This work was supported in part by The National Key Research and Development Program of China under grant 2020AAA0106000.

\bibliographystyle{ACM-Reference-Format}
\bibliography{bib_full_name}

\appendix
\section{Appendix for reproducibility}
\subsection{Search Space}
Since the hyper-parameter of the CF model can also be regarded as a part of the model from the perspective of AutoML. In our experiments, we consider learning rate and embedding size as a component of the search space. Thus, the whole search space, including the choices of user/item input encoding, user/item embedding function, interaction function, prediction function, embedding size, and learning rate. This makes there are eight one-hot encodings to represent the choice of each component.
\subsection{Search Strategy}
In our approach, we have some key hyper-parameters. For the size of the sampled model set, we set $K_1$ and $K_2$ as 10 and 10. We have also tried other settings and find the difference is minor. The stopping criterion of our approach is set to \textit{no improvement for 100 samples after beating state-of-the-art human-crafted models}.
The MLP predictor is set to be 2-layer MLP, with embedding size as \{8, 4\}, with a few parameters since the size of one-hot encodings of each model is not large.

For the training of our MLP predictor, the learning rate is set to 0.001 with an Adam optimizer, which is found effective. For the pairwise learning of the MLP predictor, we random build pairs from all of the evaluated models and use BPR loss to train the predictor.
\subsection{Performance Report}
To make the results convincing and reliable, for each experiment, we repeat them with different random seeds for three times and then calculate the average values.

\subsection{Dataset}\label{sec::appen::dataset}

\begin{table}[h]
	\small
	\centering
	\caption{The statistics of utilized datasets. (Density means the density of the user-item matrix; rating form means 1-5 ratings and interaction form means binary values.)}\label{tab:stat}
	\begin{tabular}{c|c|c|c|c|c}
		\hline
		Name     & \#User & \#Item & \#Records & Density & Form\\ \hline
		ML-100K   &  943   & 1,682  &        100,000        & 0.06304 & Rating\\ \hline
		ML-1M    &  6,040  &  3,952  &       1,000,209       & 0.04190& Rating \\ \hline
		Yelp   & 15,496  & 12,666  &        931,836         & 0.00474& Rating \\ \hline
		Amazon-Book & 11,899  & 16,196  &        911,786         & 0.00464& Rating\\ \hline
		Pinterest  & 55,187 & 9,916  &       1,500,809       & 0.00274& Interaction \\ \hline
	\end{tabular}
\end{table}

As discussed in the introduction, the datasets of different CF tasks are different. To make the experimental results more convincing, we choose five real-world raw datasets and build eight datasets for evaluation.
It is worth mentioning that for the rating prediction, the rating data, such as book-rating~\cite{ziegler2005improving}, movie-rating~\cite{bennett2007netflix}, etc., are widely used; while for item ranking tasks, the data is in the binary form, 0 for no-interaction and 1 for observed interaction. In our experiments, we convert some rating datasets to binary data for the item ranking task.
\begin{itemize}[leftmargin=*]
	\item \textbf{MovieLens-100K\footnote{https://grouplens.org/datasets/movielens/100k}.} This is a widely used movie-rating dataset containing 100,000 ratings on movies from 1 to 5. We also convert the rating form to binary form, where each entry is marked as 0 or 1, indicating
	whether the user has rated the item.
	%
	\item \textbf{MovieLens-1M\footnote{https://grouplens.org/datasets/movielens/1m}.} This is a widely used movie-rating dataset containing 1,000,000 ratings on movies from 1 to 5. Similarly, for MovieLens-100K, we also build an implicit dataset.
	\item \textbf{Yelp\footnote{https://www.yelp.com/dataset/download}.} 
	This is published officially by Yelp, a crowd-sourced review forum website where users can write their comments and reviews for various POIs, such as hotels, restaurants, etc.
	
	\item \textbf{Amazon-Book}\footnote{jmcauley.ucsd.edu/data/amazon.} 
	This is a book-rating dataset collected from users' uploaded review and rating records in Amazon. 
	\item \textbf{Pinterest\footnote{https://pinterest.com}.} This implicit data is constructed by [8] for a task of image recommendation collected from Pinterest, a popular social media website. 
\end{itemize}

\subsubsection{Preprocessing}
With a commonly used manner, we filter out users and items less than 20 records in the original datasets.
For all the pre-processed datasets, we split the data with 8:2:2 to generate training set, validation set, and test set.
For the model evaluation, we use the performance on the validation set to judge the convergence. 

\subsubsection{Implicit and Explicit}
For the first three explicit datasets, we convert them to the implicit dataset for item ranking tasks.
For every observed rating, we replace the real-value with 1, representing an observed user-item interaction in the matrix.

\subsubsection{Metrics}

\begin{itemize}[leftmargin=*]
	\item \textbf{RMSE}. Root Mean Square Error (RMSE) is the standard deviation of the residuals (prediction errors of the real scores and predicted rating scores)
	\item \textbf{MAE}. Mean Absolute Error (MAE) is the average absolute distance between predicted scores and real scores.
\end{itemize}

\begin{itemize}[leftmargin=*]
	\item \textbf{Recall} Recall@K measures the ratio of test items that have been successfully contained by the top-K item ranking list. 
	\item  \textbf{NDCG} Normalized Discounted Cumulative Gain (NDCG) complements Recall by assigning higher scores to the hits at higher positions of the ranking list.
\end{itemize}

\subsection{Detailed Settings of Baselines}
\subsubsection{Existing CF Models}
For all the single and fused models, we follow the original paper's settings. We conduct the same grid search on hyper-parameters, including learning rate and embedding size for a fair comparison.
\subsubsection{NeuMF+NASNet}
NeuMF is a fused model of GMF and MLP, and the original paper claims its MLP architecture can be carefully chosen to improve the recommendation performance.
Since GMF's architecture is quite simple, we apply a NAS method, NASNet, to search the neural architecture of MLP. To be specific, we build a controller model of which the input is a vector to denote a kind of MLP architecture. Here we use an RNN model as the controller, following the original NASNet paper. With the generated model candidates by the RNN controller, we can evaluate its performance, which is further considered as the reward of Policy Gradient to train the RNN controller.
We report the performance when it converges and then compares it with the best-searched model in our approach.

\subsection{Detailed Settings of The Implementation}
We implement all models with PyTorch\footnote{https://pytorch.org} and run all experiments on a 64-core Ubuntu 14.04 server with 8 Nvidia GeForce RTX 2080Ti GPUs and 128G memory.
It takes about 2-3 hours to search for a dataset with about 1 million interactions. 

%
\end{document}